\documentclass[prb,preprint]{revtex4-1}

\usepackage{lmodern}
\usepackage[utf8]{inputenc}
\usepackage[T1]{fontenc}

\usepackage{graphicx}
\usepackage{subfig}

\usepackage{amsmath}
\usepackage{amssymb}
\usepackage{bbm}

\newcounter{mycount}

\newcommand{\p}{\partial}
\newcommand{\zbar}{\overline{z}}

\newcommand{\id}{\mathbbm{1}}

\newcommand{\be}[1]{ \begin{eqnarray} \mbox{$\label{#1}$} }    
\newcommand{\ee}{\end{eqnarray}}

\usepackage{braket}

\begin{document}

\title{Linear dependencies between Composite Fermion states}
\author{M. L. Meyer}
\author{O. Liab\o tr\o}
\author{S. Viefers}
\affiliation{Department of Physics, University of Oslo, P.O. Box 1048 Blindern, 0316 Oslo, Norway}

\date{\today}

\begin{abstract}
It has been observed that the composite fermion (CF) approach tends to overcount the number of linearly independent candidate states for fixed sets of quantum numbers [number of particles, total angular momentum, and (pseudo)spin if applicable]. That is, CF Slater determinants that are orthogonal before projection, may lead to wave functions that are identical, or possess linear dependencies, after projection. This has been pointed out both in the context of rotating bosons in the lowest Landau level, and for excited bands of the (fermionic) fractional quantum Hall effect. We present a systematic approach that enables us to reveal \emph{all} linear dependencies between bosonic compact states in the lowest CF ``cyclotron energy'' sub-band, and almost all dependencies in higher sub-bands, at the level of the CF Slater determinants, i.e. before projection, which implies a major computational simplification. Our approach is introduced for so-called simple states of two-species rotating bosons, and then generalised to generic compact bosonic states, both one- and two-species. Some perspectives also apply to fermionic systems. The identities and linear dependencies we find, are analytically exact for ``brute force'' projection in the disk geometry. 
\end{abstract}
\pacs{ }

\maketitle


\section{Introduction}
\label{sec:intro}
Throughout the history of quantum Hall physics, a particularly successful line of research has been the construction of explicit trial many-body wave functions, notably the famous Laughlin wave function\cite{laughlin83}, the phenomenology of composite fermions\cite{jainbook}, and more recently various schemes applying to non-Abelian states\cite{moore91,read99}. The idea is that while not exact [at least not in the case of Coulomb interaction], the thus obtained wave functions capture the important topological properties of the state at hand. Many of the methods developed in the context of the quantum Hall effect have recently been applied to cold atom systems. This is of interest thanks to the impressive experimental developments in generating artificial magnetic fields in atomic Bose condensates, either by rotation\cite{roncaglia11,gemelke10} or other methods\cite{lin09, dalibardreview, julia-diaz}. The hope is to be able to realize strongly correlated states of the quantum Hall type in cold atom systems. This would provide a setting where parameters like disorder and interaction strength are controllable and tunable to a much larger degree than for electrons in semiconductor heterostructures and might, eventually, be superior to the electronic quantum Hall system for e.g. studying topological quantum computing.

As mentioned above, the composite fermion approach has been a great success in describing a large number of quantum Hall states, including their fractional excitations\cite{jainbook}. Later, this formalism was modified to successfully describe rapidly rotating Bose gases in the lowest Landau level\cite{reviews}. Most recently\cite{meyer14, jain13,grass13}, a generalised version of the composite fermion formalism, including a (pseudo)spin degree of freedom\cite{jainbook} was applied to two-species Bose condensates, which can be realized experimentally in various ways\cite{mondugno02,bloch01,hall98}. Remarkably, in the boson studies, it turned out that the CF formalism produces close to exact wave functions even at the lowest angular momenta (typically smaller the the number of particles), i.e. far outside the actual quantum Hall regime for which it was originally intended. One particular issue that was noted for low angular momenta both in the single species case\cite{viefers00,korslund,viefers10}, for two-species bosons\cite{meyer14}, and in fact also in the context of highly excited states of electronic quantum Hall states\cite{wu95,balram13}, is that the CF formalism frequently produces too many candidate states, or rather ``hidden'' linear dependencies. More precisely, the number of seemingly independent CF Slater determinants before lowest Landau level (LLL) projection obeying the pertinent physical constraints [desired total angular momentum and (pseudo)spin quantum numbers, translation invariance etc] is often considerably larger than the number of linearly independent CF states after projection.
For example, for 2+6 particles and angular momentum 4, there are 41 distinct pairs of CF Slater determinants obeying all the physical constraints. However, after performing the projection -- basically since various combinations of derivatives acting on the terms of the Jastrow factor, can conspire to result in the same polynomial -- it turns out that there are only three linearly independent ones (which coincides with the actual dimension of the relevant eigenspace)\cite{meyer14}.
This is a mathematical feature of the CF formalism that remains to be fully understood, and the present paper aims to make progress in this direction. A systematic understanding of this issue is obviously of practical interest: numerical calculations would be significantly simplified if one were able to systematically identify the set of linearly independent CF states  \textit{a priori}, from the unprojected form of the CF Slater determinants, rather than having to carry out the projection brute force, and looking for linear dependencies among the resulting  polynomials. This is particularly important in the low angular momentum regime, where LLL projection amounts to a very large number of derivatives, and is thus computationally heavy.

The main part of this paper concerns the systematic classification of linear dependencies in the CF description of two-species Bose gases with homogeneous interaction. We study the angular momentum regime $L \leq N\cdot M$ where $N$ and $M$ denote the particle numbers of the two species. In particular we focus on what we refer to as \emph{simple states}. These were identified in Ref.\onlinecite{meyer14} as a certain subset of all CF candidate states at a given number of particles and angular momentum. Technically, they are characterized by having at most one composite fermion occupying each $\Lambda$-level in the CF Slater determinants, and they can be shown to minimize the $\Lambda$-level ``cyclotron'' energy of the state. While diagonalisation within the full set of CF candidate states recovers basically the entire yrast spectrum for low angular momenta exactly or near exactly, the simple states still give a very accurate description of the low-lying part of the yrast spectrum, with typical overlaps $>97\%$ for 12 particles\cite{meyer14}. Even when restricting to simple states, the number of seemingly different CF candidates still tends to get vastly larger than the number of linearly independent states after projection. 
The reason for focussing on simple states first is that the concepts and techniques we use to systematically reveal linear dependencies are most easily introduced this way. However, many of these ideas also apply more generally to non-simple states (including single-species), as well as fermionic systems. These cases will be addressed later in the paper.

To summarize, our goal is to develop a systematic way of revealing linear dependencies at the level of the Slater determinants themselves, so that a minimal set of linearly independent CF basis states is identified before explicitly performing LLL projection. Our approach involves a letter string notation to represent the occupation patterns of the CF Slater determinants. Exploiting the fact that all so-called compact states (at some given angular momentum $L$) are translationally invariant leads to linear relations between states at $L-1$. Further identities are found from a generalised version of translation invariance, as well as various ways of reordering the occupation patterns of the Slater determinants. All identities are proven analytically and supported by numerical calculations. 
Since this part of the paper is rather technical, we we will illustrate with examples, and defer some details of the proofs to an Appendix.
We start by summarizing some necessary background theory in section \ref{sec:theory}. Sec \ref{sec:simple} treats in detail the case of simple states, while general compact states and the fermionic case are discussed in sections \ref{sec:compact} and \ref{sec:fermi}, respectively. We end with summary and future perspectives in section \ref{sec:concl}.


\section{Two-component rotating Bose gases}
\label{sec:theory}
We here give a quick summary of the model for two-species Bose gases in the lowest Landau level with homogeneous interaction, including their description in terms of composite fermions. For a more detailed introduction we refer the reader to Ref.\onlinecite{meyer14}. The Hamiltonian for our system, two species of bosons in a two-dimensional harmonic trap of strength $\omega$, rotating at frequency $\Omega$, is
\begin{equation}
H = \sum_{i=1}^{N+M} \left( \frac{\mathbf{p}_i^2}{2m}+\frac{1}{2}m\omega^2\mathbf{r}_i^2 - \Omega l_i \right)
   + \sum_{i<j=1}^{N+M}2\pi g \delta(\mathbf{r}_i - \mathbf{r}_j).  
\end{equation}
Here $M$ denotes the number of particles of the majority species, and $N$ is the number of particles of the minority species. The single-particle angular momenta are denoted by $l_i$. We have assumed that all particles have the same mass $m$, and that the strength of the contact interaction, $g$, is independent of species. This is what we refer to as homogeneous interaction. In the weak interaction (dilute) limit this reduces to the well known lowest Landau level problem\cite{reviews} in the effective magnetic field $2m\omega$,
\begin{align}
H &= \sum_{i=1}^{N+M}(\omega-\Omega)l_i + 2\pi g\sum_{i<j=1}^{N+M} \delta(\eta_i - \eta_j).
\end{align}
In the ideal limit $(\omega - \Omega) \rightarrow 0$ one gets flat Landau levels, and so all the many-body physics of the system is determined by the interaction.
Here $\eta_j = x_j + i y_j$ are the dimensionless complex positions of the particles in units of the ``magnetic'' length $\sqrt{\hbar/(2m\omega)}$. The second sum runs over both species since the interaction is homogeneous.
Working in symmetric gauge, the lowest Landau single-body eigenstates with angular momentum $l$ are 
\begin{equation}
\psi_{0,l}(z) = N_{l} z^l \exp{(-z\bar{z}/4)} \qquad l\geq 0 \label{spwf}
\end{equation}
where we will suppress the Gaussian factor for simplicity from now on. A generic many-body wave function with fixed total angular momentum $L$ is then a homogeneous polynomial of degree $L$, symmetric in the coordinates of each species separately. As previously\cite{reviews, meyer14} we will focus on translationally invariant states, i.e. polynomials invariant under a simultaneous, constant shift ($K$) of all coordinates,
\begin{equation}
 \Psi(z+K,w+K) = \Psi(z,w).
\end{equation}
where $z$ and $w$ denote the sets of coordinates of the minority and majority species, respectively.

Due to the species-independent interaction strength, the Hamiltonian is invariant under change of species, which implies a pseudospin-1/2 invariance.
In Ref.\onlinecite{meyer14} we explained how this spin analogy can be exploited to greatly simplify the analysis of the system's many-body spectra. This aspect is of less importance in the present paper, where we will focus on the mathematical properties of CF states that are already known from Ref.\onlinecite{meyer14} to be good candidates for the low-lying states of the system.

A generic CF trial wave function for the bosonic two-species system is of the form\cite{meyer14}
\begin{equation}
 \Psi_{CF}=\mathcal{P}_{LLL} \left(\Phi_Z \Phi_W J(z,w) \right)
 \label{2swf}
\end{equation}
where $\Phi_Z$, $\Phi_W$ are Slater determinants for each species of the non-interacting CFs. They consist of the single-particle states
\begin{equation}
\psi_{n,m}(z) = N_{n,m} z^m L_n^m\left( \frac{z\bar{z}}{2} \right),
\quad m \geq -n 
\label{spwf2},
\end{equation}
where $L_n^m$ is the associated Laguerre polynomial, and $N_{n,m}$ is a normalization factor.
$J$ is a full Jastrow factor involving both species,
\be{}
J(z,w)&=&\prod_{i<j=1}^{N+M}(\eta_i - \eta_j)  \nonumber \\
	 &=&\prod_{i<j=1}^N(z_i-z_j)\prod_{k<l=1}^M(w_k-w_l)\prod_{i,k=1}^{N,M}(z_i-w_k).
\ee
$\mathcal{P}_{LLL} $ denotes projection to the lowest Landau level. The original, most direct projection method amounts to replacing the conjugate variables $\overline{z}_i$, $\overline{w}_k$ by $\partial_{z_i}$, $\partial_{w_k}$ after moving them all the way to the left in the final polynomial \cite{jainbook}. While other projection techniques are frequently employed in the literature in order to make numerical calculations less heavy, all our calculations are done using this original method. While the use of e.g. Jain-Kamilla projection\cite{jainbook} has been shown to make very little qualitative and quantitative difference in numerical calculations, our exact analytical results would presumably only be near-exact with this projection technique. We will briefly return to this point at the end of the paper.


\section{Linear dependencies for simple states}
\label{sec:simple}

\begin{figure}
\includegraphics[width=\columnwidth]{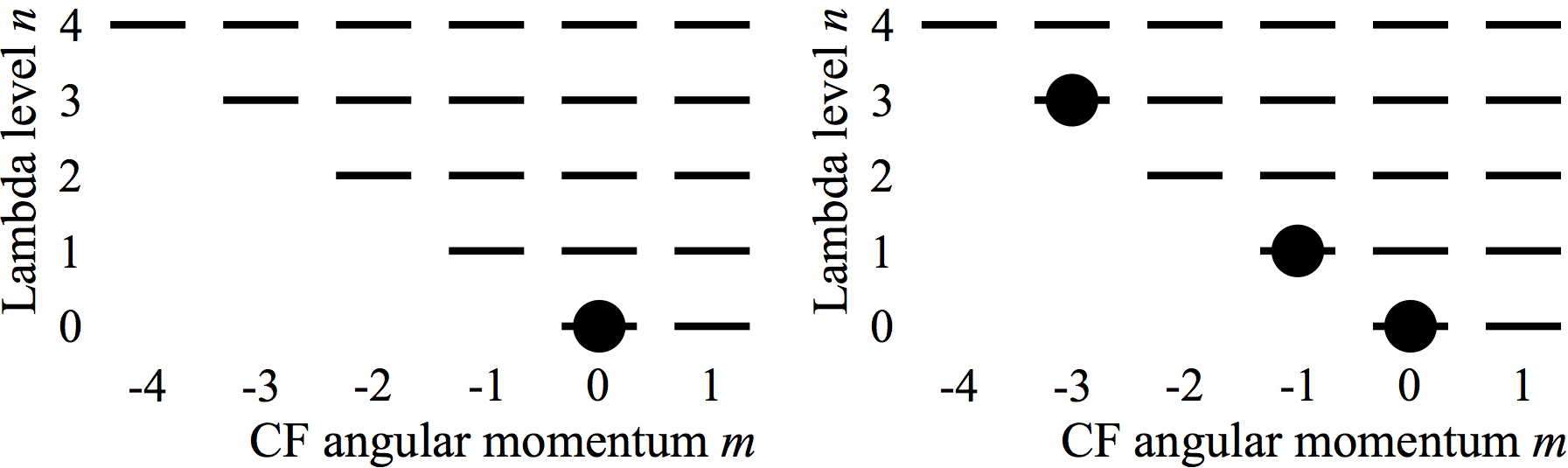}
\caption{Sketch of CF Landau (or ``$\Lambda$''-) level occupancies for the pair of Slater determinants of a simple state. Such states are characterized by at most one composite fermion of a given species occupying any Lambda level.
\label{fig:simplestate}}
\end{figure}

In this rather technical section we introduce, step by step, how to reveal linear dependencies between CF candidates at given particle numbers, total angular momentum and pseudospin quantum numbers. Each step introduces new linear relations between states, finally leading to a systematic algorithm to reduce the set of CF candidates to a basis. This section focuses on ``simple'' states, i.e. states with at most one composite fermion of each species occupying
a given $\Lambda$-level $n$, with minimal angular momentum ($m=-n$). The latter ensures translational invariance of these states. An example is given in FIG. \ref{fig:simplestate}. The polynomial part of the corresponding single-particle eigenfunctions is $\eta_{n,-n}=\bar z^n$ (neglecting the normalization constant) which translates to $\eta_{n,-n}=\partial_z^n$
after projection.

\begin{figure}
\includegraphics[width=\columnwidth]{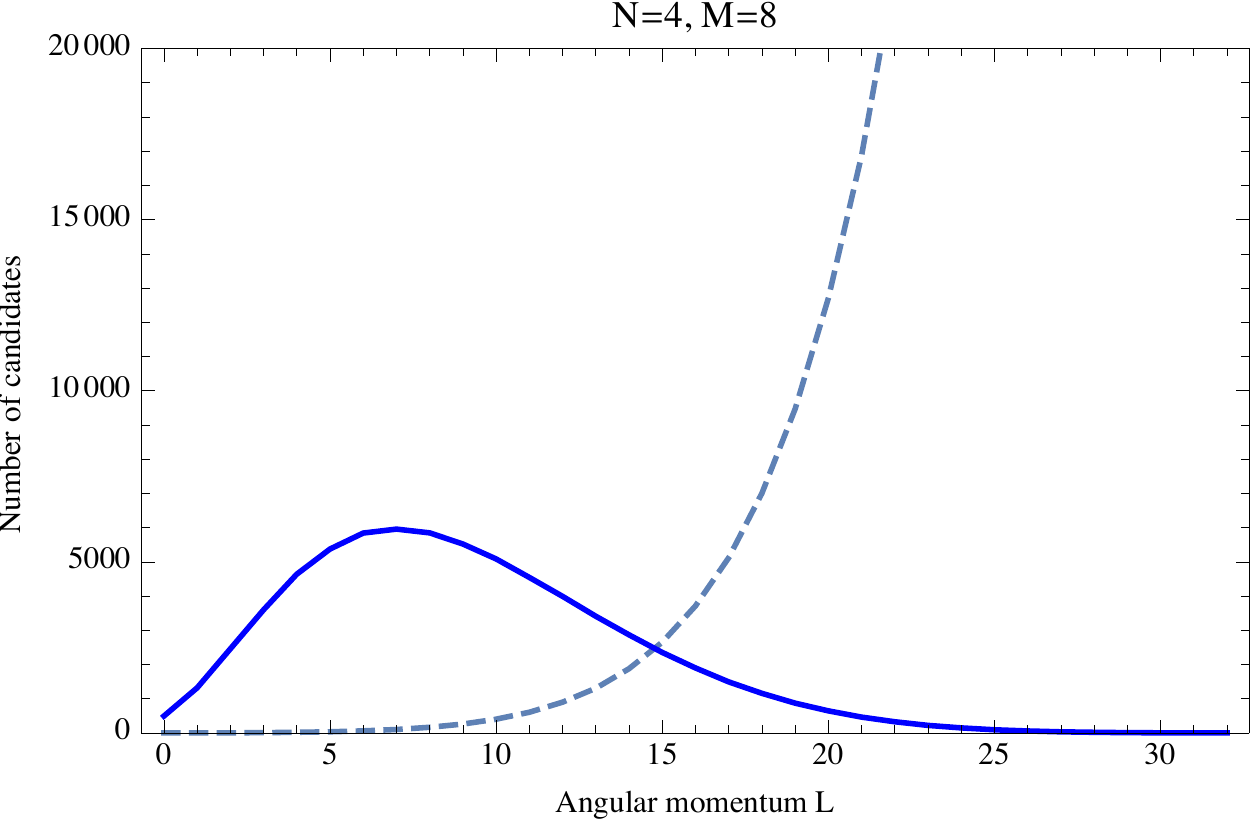}
\caption{ The solid line shows the number of naively independent simple CF candidates as function of angular momentum for 4+8 particles. The dashed line shows the number of linearly independent wave functions in the lowest Landau level.
\label{fig:simple-naive-plot}
}
\end{figure}

The possibility of having, and necessity of understanding, dependencies between simple CF candidate wave functions is apparent from FIG. \ref{fig:simple-naive-plot}. Indeed, for low enough angular momentum ($L<15$ in the case of 4+8 particles), the number of simple candidates exceeds the dimension of the LLL basis, meaning that they \emph{cannot possibly} form a linearly independent set. Based on this we infer that there exist dependencies between the seemingly independent candidates at these low angular momenta, and indeed this is also the case for higher values of $L$, as can be checked by performing the projection to the LLL. Understanding these dependencies is therefore necessary in order to efficiently use the CF construction to study the two-component Bose system.

\subsection{Letter string notation}
The simple CF wave functions form a very restricted set, which makes it possible to represent them compactly using strings of letters.  For a state with $N$ particles of type $Z$ and $M$ particles of type $W$, the single particle states will be on the form $(n,m)=(x,-x)$ with $x\in \{0,\ ..., N+M-1\}$.
Higher values of $x$ are excluded as the power of the corresponding derivative, $\partial^x$, would exceed the highest possible power of any variable in the Jastrow factor. 
Since no single-particle level can be occupied by more than one CF of a given species, there are four possibilities: A state $(i,-i)$ can be occupied by a $Z$, a $W$, both a $Z$ and a $W$, or neither a $Z$ nor a $W$. These scenarios are denoted by ``$Z$'', ``$W$'', ``$P$'' (``pair'') and ``$H$'' (``hole''), respectively. A many-body simple state is then represented by the corresponding string of $N+M$ zero-indexed letters, ordered by increasing $n$.
For example, the wavefunction for the state in FIG. \ref{fig:simplestate} (with the left figure representing $Z$-particles, and the right part corresponding to $W$-particles) can be represented as 

\be{}
\Psi(\{ z_i\}, \{ w_i\}) = 
\begin{vmatrix}
\p^0_{z_1}
\end{vmatrix}
\cdot
\begin{vmatrix}
 \p^0_{w_1} \, \p^0_{w_2} \,\p^0_{w_3}\\
\p^1_{w_1} \, \p^1_{w_2} \,\p^1_{w_3} \\
\p^3_{w_1} \, \p^3_{w_2} \, \p^3_{w_3}  \\
\end{vmatrix}
\cdot
\, J(z,w)\equiv PWHW.
\label{simple_example}
\ee

The terms in the Jastrow factor contain exactly $i$ variables of order at least $N+M-i$.  A non-zero CF wave function must therefore have at most $i$ differentiation operators of order $N+M-i$ or higher.  This implies that no suffix can contain more $P$'s than $H$'s in the letter string representation.  Equivalently, no prefix can contain more $H$'s than $P$'s.

\subsection{Translation invariance}

Simple states are compact and thus translation invariant\cite{jainbook}, in the sense that their polynomial part satisfies
\be{}
\Psi(\{ z_i\}, \{ w_i\}) = \Psi(\{ z_i+K\}, \{ w_i+K\}).
\ee
We define the differentiation operators
\be{}
\Delta_Z\equiv\sum_{i=1}^N \partial_{z_i},\ \ \Delta_W\equiv\sum_{j=1}^N \partial_{w_j}. 
\ee
Now, an equivalent way of stating translation invariance is
\be{}
(\Delta_Z+\Delta_W)\Psi=0.
\ee
These differentiation operators act on the CF polynomials by raising single particle states $(i,-i)$ to $(i+1, -i-1)$.  For the polynomial $PWHW$, this gives
\be{}\label{ola1}
(\Delta_Z+\Delta_W)PWHW=WPHW+PHWW=0.
\ee
Applying translation invariance on states of angular momentum $L$ thus gives linear dependence relations for states of angular momentum $L-1$. Other examples are
\be{}
(\Delta_Z+\Delta_W)PPWZHH=PWPZHH+PPWHZH+PPHPHH=0
\ee
and
\be{}
(\Delta_Z+\Delta_W)PWZWH=WPZWH+PWHPH+PHPWH+PWZHW=0.
\ee
These dependence relations reduce the number of candidate states, but not generally to a linearly independent basis set, except in some special cases when the total number of particles is low.

\subsection{Blocks}
\label{sec:blocks}
A further reduction of the number of CF candidates, explored in this subsection, is due to invariance of the final CF polynomial under certain ways of permuting the occupation patterns of the Slater determinants.
As mentioned, a string will represent the zero polynomial if it has a prefix with more $H$'s than $P$'s and is therefore not interesting.  An interesting situation occurs, however, when a prefix has an equal number of $P$'s and $H$'s.  If the prefix is of minimal, positive length we call it a \emph{block}. We repeat this with the remainder of the string until the whole string is partitioned into blocks. We write such a partitioned string as
\be{}
X_1 \circ X_2 \circ \ldots \circ X_n,
\ee
where the $X_i$ are blocks, and $\circ$ denotes concatenation of strings. For example, the resulting polynomials in equation (\ref{ola1}) have the blocks
\be{}
WPHW = W \circ PH \circ W,\ \ PHWW = PH \circ W \circ W.
\ee
This partitioning of the string is useful because applying the Slater determinants to the Jastrow factor will only give non-zero terms if the differentiation operators from the first block are applied to the variables with the lowest exponents in the Jastrow factor. The differentiation operators from the second block  are then applied to the lowest remaining exponents, and so on. This is used in Appendix \ref{sec:app} to show that the corresponding polynomial is invariant under permutations of blocks up to a sign.  The commutation rule is
\be{}
X_1 \circ X_2 = (-1)^{N_1\cdot M_2+N_2\cdot M_1} X_2 \circ X_1 \label{commutation-rule},
\ee  
where the $X_i$ are blocks, $N_i$ denotes numbers of $Z$ particles, and $M_i$ numbers of $W$ particles in block $X_i$. Incidentally, the linear dependence relation (\ref{ola1}) can now alternatively be viewed as a consequence of the permutation rules for blocks, since 
\be{}
WPH = W \circ PH = (-1)^{0\cdot1+1\cdot1}PH \circ W = -PHW.
\ee

Pure permutation invariance is not the only useful consequence of blocks; they also allow us to expand the concept of translation invariance. We note that, up to a combinatorial factor, the polynomial of a CF state with multiple blocks equals the symmetrization of the polynomials of the blocks when they are considered as individual states.  Since each block is translation invariant, we can differentiate the blocks separately and thus obtain more dependence relations.  For example, since
\be{}
(\Delta_Z+\Delta_W)PWZH=(\Delta_Z+\Delta_W)PWH=0,
\ee
the following dependence relations can be derived from the state $PWZHPWH$:
\be{}
(PWHZ+PHPH) \circ PWH=PWHZPWH+PHPHPWH=0,\\
PWZH \circ (WPH+PHW)=PWZHWPH+PWZHPHW=0.
\ee
In contrast, using translation invariance of the whole state would merely give
\be{}
PWHZPWH+PHPHPWH+PWZHWPH+PWZHPHW=0.
\ee
In general, translation invariance for individual blocks means that if we have a dependence relation
\be{}
\sum_i X_i=0,
\ee{}
where the $X_i$ are letter string representations of wave functions, then it is also true that
\be{}
\sum_iX_a \circ X_i\circ X_b=0,
\ee
where $X_a$, $X_b$ are arbitrary letter string representations of wave functions.

\subsection{Reflection}
In this section we prove that CF polynomials of simple states are invariant under what we refer to as reflection symmetry.  This symmetry is somewhat related to translation invariance, but it is not captured in the dependence relations found above. Reflection is defined as follows. Assume $X$ is a letter string representing a CF wave function or block. Then we define the reflected string $X'$ as the result of the following operations:
\begin{enumerate}
\item Invert $X$, i.e. $ABCD\rightarrow DCBA$.
\item Swap $H\leftrightarrow P$. 
\end{enumerate}
Some examples are
\be{}
PWPZHH'=PPZHWH,\ \ PPWZPHZWHH'=PPWZPHZWHH.
\ee
Counting from the beginning of the reflected state, the $Z(W)$ particles occupy the positions that were vacant of $W(Z)$ particles in the original state when counting backwards from the end.  It is not difficult to see that reflection leaves $N, M$ and $L$ invariant, and that $X''=X$.  The non-trivial result is that 
\be{}\label{reflectionsymmetrytheorem}
X'=(-1)^{NM+L}X.
\ee   
We need a small lemma for the proof:\\

\be{}\label{reflectionlemma}
(\Delta_ZX)'=\Delta_W(X')
\ee
Applying $\Delta_Z$ and then reflection to $X$ can be described by the list of operations 

\begin{enumerate}
\item Create the sum over every possible $Z$ moving to the right.
\item Invert X, i.e. $ABCD\rightarrow DCBA$.
\item Swap $H\leftrightarrow P$. 
\end{enumerate}
Or, equivalently
\begin{enumerate}
\item Invert X.
\item Create the sum over every possible $Z$ moving to the \emph{left}.
\item Swap $H\leftrightarrow P$, or equivalently, $(\text{missing } Z) \rightarrow W$, $(\text{missing } W) \rightarrow Z$.
\end{enumerate}
Further, since a $Z$ moving to the left is the same as a missing $Z$ moving to the right, we get that this procedure is also equivalent to
\begin{enumerate}
\item Invert X.
\item Swap $H\leftrightarrow P$. 
\item Create the sum over every possible $W$ moving to the \emph{right}.
\end{enumerate}
and this final list is just the procedure that describes $\Delta_W(X')$.

We are now ready to prove reflection symmetry (\ref{reflectionsymmetrytheorem}) by induction on $L$: \\
For $L=0$, every possible string representation will consist of $N$ $Z$'s and $M$ $W$'s in some order.  Since there are no $P$'s or $H$'s, the reflected string will just be the inverse string.  We know from section \ref{sec:blocks} that $W$ and $Z$ blocks anticommute while $Z$ commutes with $Z$ and $W$ commutes with $W$.  Since one possible way of commuting the blocks so that $X\rightarrow X'$ involves moving every block past every other block, we get a sign of exactly $(-1)^{NM}$ from interchanging neighbouring $W$'s and $Z$'s.  This shows that the result is satisfied for $L=0$.\\
Now, assume that (\ref{reflectionsymmetrytheorem}) holds for $L-1$ and consider the following expression,
\be{}
\Delta_Z(X-(-1)^{NM+L}X')=\Delta_Z X - (-1)^{NM+L} \Delta_Z (X') \nonumber \\
=\Delta_ZX+(-1)^{NM+L}\Delta_W(X')=(\Delta_Z X) - (-1)^{NM+(L-1)}(\Delta_Z X)'=0. \label{zder}
\ee 
The second equality is due to translation invariance, $\Delta_Z X=-\Delta_W X$, then we use lemma (\ref{reflectionlemma}) and finally the induction hypothesis for $L-1$, since $\Delta_Z X$ is a sum of strings with angular momenta $L-1$. Since $\Delta_Z = - \Delta_W$, we also have
\be{}
\Delta_W(X-(-1)^{NM+L}X')=0. \label{wder}
\ee 

Since $X-(-1)^{NM+L}X'$ is a symmetric function in both $z$ variables and $w$ variables, Eqs. (\ref{zder}-\ref{wder}) imply that $X-(-1)^{NM+L}X'$ is constant in all its variables. Since $L>0$ it must vanish. Thus, (\ref{reflectionsymmetrytheorem}) holds for $L$, which completes the proof.

\subsection{Generalised translation invariance}

Applying the results of block permutation, reflection symmetry and translation invariance does reduce the number of CF candidate states significantly, but still leaves dependencies that cannot be explained from these symmetries.  The final piece to the puzzle is a generalised version of translation invariance.
We define differentiation operators
\be{}
\Delta_{Za} \equiv \sum_{i=1}^N \frac{\partial^a}{\partial z^a},\ \ \Delta_{Wa} \equiv \sum_{j=1}^M\frac{\partial^a}{\partial w^a}, \quad a \in \mathbb{N}.
\ee

These operators commute with the Slater determinants, since the latter consist only of derivatives for simple states. They can therefore be applied directly onto the Jastrow factor in order to study their action on the simple states. Since the Jastrow factor is anti-symmetric in all its variables, and $\Delta_{Zn}+\Delta_{Wn}$ is symmetric in the same variables, the product $(\Delta_{Zn}+\Delta_{Wn})J(z,w)$ is anti-symmetric too.  However, since the Jastrow factor contains all powers from $0$ to $N+M-1$, any non-zero resulting term from the differentiation must have two exponents that are equal. The anti-symmetry then ensures that the result is zero.

The simple states that result from this higher order differentiation can be obtained in the same way as for translation invariance, using letter string notation. Applying $\Delta_{Za}$ to a string results in a sum over every possible move of a $Z$ $a$ places to the right. A difference from translation invariance is that such a move can cause a $Z$ to move past another $Z$, and this causes a minus sign. The reason is that the polynomials that the strings represent are defined with differentiation powers occurring in increasing order in the Slater determinants. For example, we have
\be{}
(\Delta_{Z2}+\Delta_{W2})PPWWZHH &=& -WPPWZHH + PWWPZHH - PPHWPHH+ PPWHZWH \nonumber \\
  &=&0.
\ee 
As with regular translation invariance, $(\Delta_{Za} + \Delta_{Wa})$ may also be applied to individual blocks to create one dependence equation per block in a state. In addition to producing more dependence equations, this approach has the added benefit of constraining the values of $a$ that need consideration. To see this, consider a string composed of blocks
\be{}
X_1 \circ \ldots \circ X_i \circ X_{i+1} \circ \ldots \circ X_n.
\ee
If the length of the block $X_i$ is $l_i$, the application of, say, $\Delta_{Za}$ on this block will necessarily move a $Z$ from $X_i$ to $X_{i+1}$ if $a \geq l_i$. Such a move will either turn a $P$ into a $W$, or a $Z$ into an $H$ in $X_i$. But then, the prefix $X_1 \circ \ldots \circ X_i$ contains more $H$'s than $P$'s, and consequently, the string represents the zero polynomial. The conclusion is that any term where a $Z$ or $W$ moves from one block to another in the original string results in a vanishing polynomial. In particular, when acting with $\Delta_{Za}+\Delta_{Wa}$ on a block $X_i$ with length $l_i$, only $a<l_i$ will give something non-zero.

\subsection{Combined results}
\label{sec:comb-res-simple}

We are now in a position to describe the main results of this work. In this section we present an algorithm that reduces the number of CF candidates for a given $N,M,L$ by as much as possible, according to the known linear dependence relations detailed above. The algorithm is easy to implement because the set of all simple states is closed under the operations leading to dependence relations.

The algorithm can be summarized in the following steps:
\begin{enumerate}
\item Construct all simple CF candidates at $N,M,L$. This set of candidates is named $\left\{ \Psi_{CF} \right\}$.
\item Construct dependence relations by applying all combinations of permutation and reflection of blocks. Use the dependence relations to reduce the set $\left\{ \Psi_{CF} \right\}$.
\item For all $\Delta L = 1,\ldots, N+M-1$, do:
\begin{enumerate}
\item Construct all simple CF candidates at $N,M,L+\Delta L$.
\item Construct additional dependence relations using the generalized translation invariance condition (\ref{gen-trans-comp}) with $a=\Delta L$, applied to all the blocks of length greater than $a$ of the states in step 3(a). Simplify the new dependence relations using those found in step 2.
\end{enumerate}
\item Use the relations found in step 3 to further reduce the set $\left\{ \Psi_{CF} \right\}$.
\end{enumerate}

Using the algorithm presented above, we can compare the size of the reduced $\left\{ \Psi_{CF} \right\}$ to the actual number of linearly independent wave functions, found by a brute-force calculation. We find that $\left\{ \Psi_{CF} \right\}$ is fully reduced in all cases, i.e. that the reduced $\left\{ \Psi_{CF} \right\}$ is a linearly independent set of wave functions, and thus constitutes a basis for the simple CF candidates. The algorithm thus succeeds in removing all linear dependencies without doing any projections to the LLL, i.e. without actually comparing the very complicated polynomials in the final wave functions. The statements above have been verified numerically for up to 14 particles.

\begin{figure}
 \subfloat[]{
  \includegraphics[width=0.8\columnwidth]{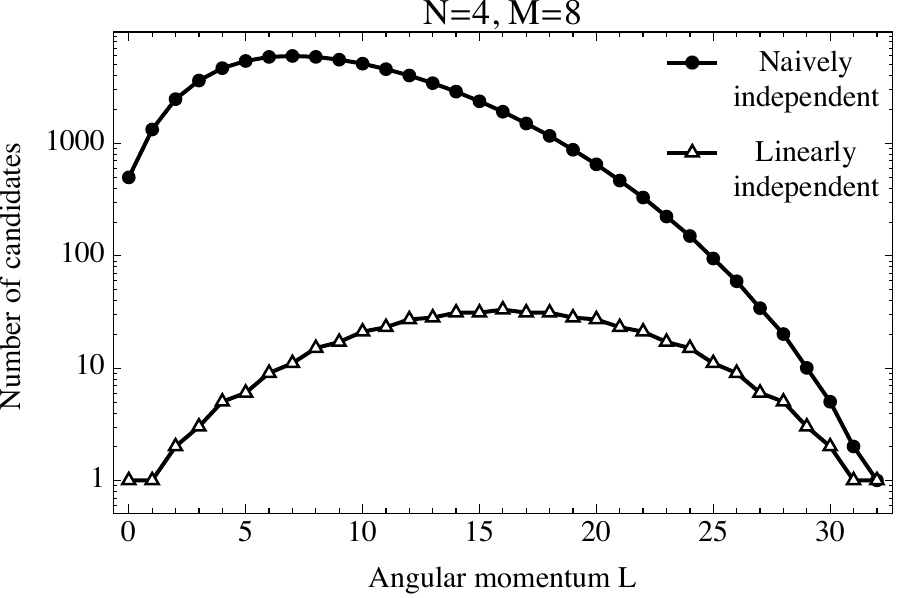}
\label{plot:simple-number}
 }\\
 \subfloat[]{
  \includegraphics[width=0.8\columnwidth]{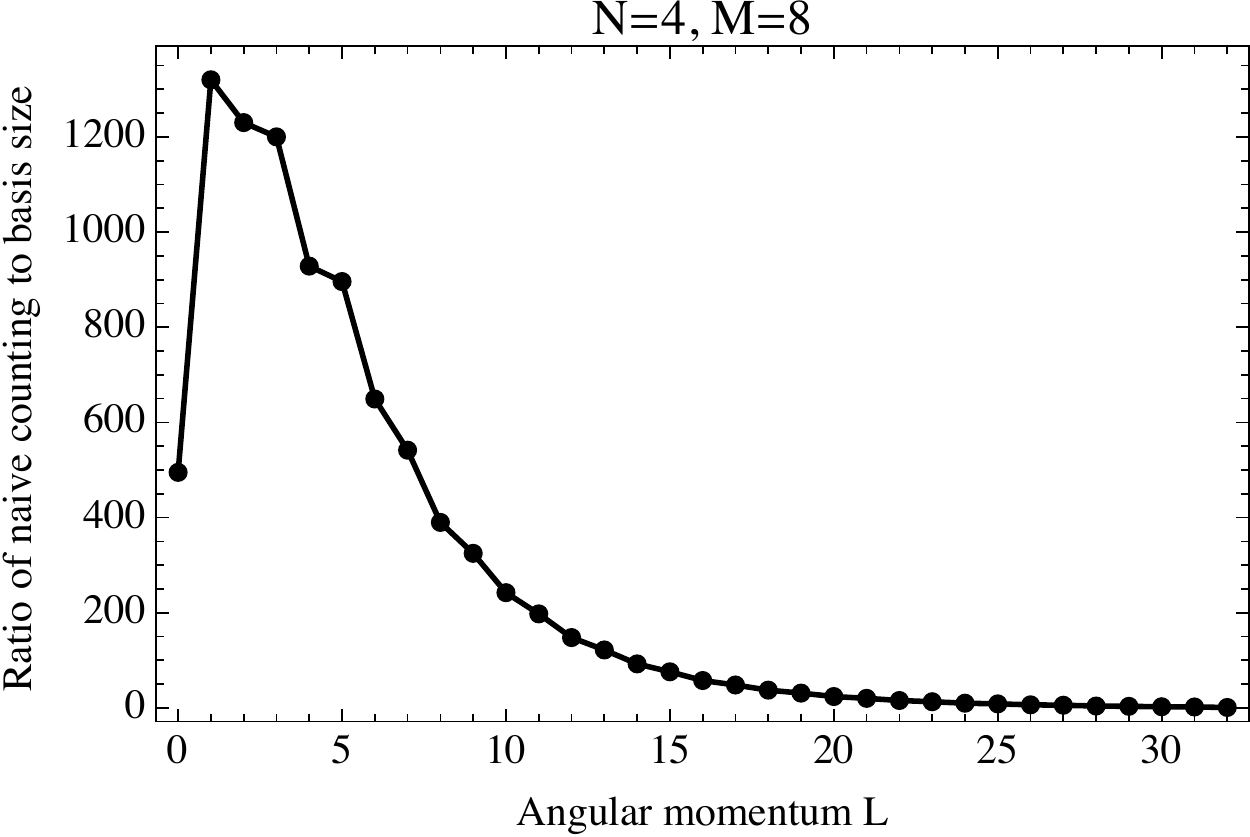}
\label{plot:simple-ratio}
 }
\caption{(a) Logarithmic plot of the number of naively independent, and actually independent, simple CF candidates, for 4+8 particles. (b) The corresponding ratio of the number of naively independent to linearly independent simple CF candidates.}
\end{figure}

To illustrate the benefit of applying the algorithm above, FIG. \ref{plot:simple-number} shows the number of naively independent, and actually independent, simple CF candidates, while FIG. \ref{plot:simple-ratio} show their ratio, for 4+8 particles. For low $L$ the size of the basis is two to three orders of magnitude smaller than the size of the naive set of candidates, meaning that projecting all of them would be extremely wasteful, computationally. For somewhat larger $L$, the ratio decays rapidly with $L$, but the absolute number of states is still significantly decreased by the algorithm.

\begin{figure}
\includegraphics[width=0.8\columnwidth]{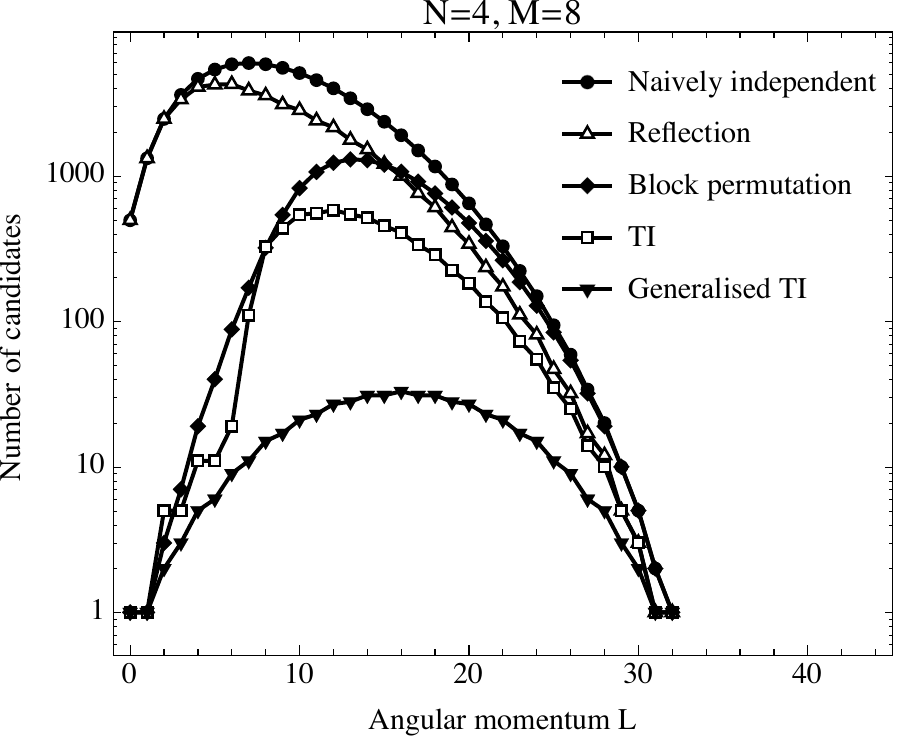}
\caption{Logarithmic plot showing the number of seemingly independent simple states after reducing the naive set by the mechanisms described in this section, again for 4+8 particles. TI stands for translation invariance.
\label{fig:reductions}}
\end{figure}

We can see how reflection symmetry, block permutation, ordinary and generalised translation invariance contribute to the explanation of linear dependencies. FIG. \ref{fig:reductions} shows the number of candidates after removing dependencies caused by the different mechanisms. First, we notice that overall, reflection seems to contribute the least, particularly for small $L$. This is easy to explain: reflection removes one out of two blocks that are reflections of each other, but only for blocks that are not reflected unto themselves ($PPWHH$ is an example of a block equal to its reflection). For low $L$, many blocks are short (contributing no or little angular momentum to the total), and given that the shortest non-self-reflected blocks are the four-particle blocks $PZWH$ and $PWZH$, reflection provides few dependence relations. Second, block permutation is increasingly important with decreasing $L$ for the same reason: as $L$ decreases we can form more and more small blocks, increasing the number of candidates that only differ by the order of the blocks. On the other hand, the largest $L$ for which one can have more than one block is $L = N(M-1)$. This is the two-block state
\be{}
\underbrace{P\ldots P}_N \ \underbrace{W\ldots W}_{M-N-1} \ \underbrace{H\ldots H}_N \circ W
\ee
For $L > N(M-1)$, all candidates are one-block states, and permutation symmetry tells us nothing.

Last, we notice that the number of candidates after generalised translation invariance is symmetric about $L_{mid} = N M/2$. In fact, the generalised TI-line equals the fully reduced line from FIG. \ref{plot:simple-number}, and this is also the case for the other numbers of particles we have studied numerically. However, remembering that reflection and block permutation can be exploited very easily \emph{before} applying the generalised translation invariance conditions, this is a valuable, and often necessary, thing to do: especially for small $L$, neglecting to exploit block permutation will lead to extremely many unnecessary generalised TI-equations. The symmetry about $L_{mid}$ lends hope to the idea that there might exist an analytic combinatoric formula for the dimension of the simple CF basis at any $N,M,L$. We comment on this in section \ref{sec:concl}.


\section{General compact CF states}
\label{sec:compact}

The simple state with the highest possible angular momentum $L$ for given $N,M$ is produced by filling the ``ladder'' of CF orbitals $\{(n,-n)\}$ from below, giving the letter string representation:
\be{}
\underbrace{P\ldots P}_N \ \underbrace{W\ldots W}_{M-N} \ \underbrace{H\ldots H}_N
\ee{}
This state has angular momentum $L=N \cdot M$. For $L>N \cdot M$, no simple states exist. The relevant CF candidates for the low-lying eigenstates with angular momentum $L>N \cdot M$ are known as \emph{compact states} \cite{jain-kawamura}. A single species CF candidate is said to be compact if its Slater determinant is such that:
\begin{enumerate}
\item for any occupied CF orbital $\psi_{n,m}$, $m>-n$, the orbital $\psi_{n,m-1}$ is also occupied, and
\item for any $\Lambda$-level $n$ occupied by $k$ CFs, the level $n+1$ is at most occupied by $k+1$ CFs.
\end{enumerate}
As mentioned earlier on, simple states satisfy these requirements trivially, i.e. simple states are special cases of compact states. Compact wave functions have some well-known properties\cite{jainbook}.
First, compact states obey translation invariance of the polynomial part of the wave function, like simple states do. Second, one can use row reduction on the Slater determinants after projection to show that the effective CF orbitals can be written\cite{jainbook}

\be{}
\psi_{n,m}(z_i) = z_i^{m+n} \partial_{z_i}^n \label{sps-compact-1}
\ee

For a two-component compact CF candidate, the Slater determinants of each species are required to be compact separately. The compact state with the largest obtainable $L$ has all particles sitting compactly in the lowest $\Lambda$-level in both Slater determinants, so this state is nothing but the ``221'' state, 
$\psi_{221} = \prod_{i<j} (z_i - z_j)^2 \prod_{k<l} (w_k - w_l)^2 \prod_{m<n} (z_m - w_n)$. 
The angular momentum of this state is $L_{221} = N(N-1) + M(M-1) + N \cdot M$. The compact states are thus our objects of interest in the angular momentum range $N\cdot M < L \leq L_{221}$.

We now consider the generalizations of the linear dependence concepts presented for simple states to general compact states. Apart from certain modifications which will be described, all the concepts developed for simple states will also apply to compact states, except for reflection symmetry, for which we have not found a suitable generalization. The modifications are necessary due to the fact that the set of compact states is \emph{not} closed under the action of (generalised) translation invariance or permutation of blocks. These modifications will be the topic of this section.

\subsection{Notation}

The letter string notation is not adequate to describe compact states. In this section, we will, in addition to simply listing which CF orbitals are occupied or writing down the Slater determinants, employ a diagrammatic representation as follows. The letters $P$, $Z$ and $W$ will signify occupation by both species, only the $Z$ species, and only the $W$ species respectively, and the possible orbitals are shown in a $m$ vs. $n$ diagram. Unoccupied orbitals will be displayed as dashes. For instance, the CF candidate which has $\psi_{2,-2}$ and $\psi_{3,-3}$ occupied by the $Z$ species, and $\psi_{0,0}$, $\psi_{0,1}$, $\psi_{2,-2}$ and $\psi_{4,-4}$ occupied by the $W$ species will be represented by the $\Lambda$-level diagram
\be{}
\begin{array}{ccccccccc}
 & 5 & - & - & - & - & - & - & -\\
 & 4 & & W & - & - & - & - & -\\
 & 3 & & & Z & - & - & - & -\\
n & 2 &  & & & P & - & - & -\\
 & 1 & & &  & & - & - & -\\
 & 0 & &  &  &  & & W & W\\
 &  & -5 & -4 & -3 & -2 & -1 & 0 & 1 \\
 &  &  &  & m
 \end{array} \label{diagram-example}
\ee
All allowed $\Lambda$-levels $n=0,\ldots,N+M-1$ are included in the diagrams for completeness, making it easier to compare candidate states, but they may be truncated after the largest value of $m$.

\subsection{Generalised translation invariance}
We wish to exploit that the Jastrow factor vanishes under the generalised translation operator:
\be{}
\left( \Delta_{Za} + \Delta_{Wa} \right)J = \left( \sum_{i=1}^N \p_{z_i}^a + \sum_{j=1}^M \p_{w_j}^a \right) J = 0 \quad \forall \ a \in \mathbb{N} \label{jas-gen-trans}
\ee
The product of one projected Slater determinant, say for the $Z$ species, and the $Z$-part of the generalised translation operator, is
\be{}
\begin{split}
 & \left( \sum_{\sigma} (-1)^{| \sigma |} \prod_{k=1}^N \psi_{n_{\sigma(k)},m_{\sigma(k)}}(z_k) \right)
\left( \sum_{i=1}^N \p_{z_i}^a \right) \\
= & \left( \sum_{\sigma} (-1)^{| \sigma |} \prod_{k=1}^N z_k^{n_{\sigma(k)} + m_{\sigma(k)}} \p_{z_k}^{n_{\sigma(k)}} \right)
\left( \sum_{i=1}^N \p_{z_i}^a \right) \\
= & \sum_{i=1}^N 
\left( \sum_{\sigma} (-1)^{| \sigma |} z_i^{(n_{\sigma(i)}+a) + (m_{\sigma(i)} - a)}
\p_{z_i}^{n_{\sigma(i)}+a}
 \prod_{k \neq i}^N z_k^{n_{\sigma(k)} + m_{\sigma(k)}} \p_{z_k}^{n_{\sigma(k)}} \right) \\
= & \sum_{i=1}^N 
\left( \sum_{\sigma} (-1)^{| \sigma |}
\psi_{n_{\sigma(i)}+a,m_{\sigma(i)}-a}(z_i)
 \prod_{k \neq i}^N \psi_{n_{\sigma(k)},m_{\sigma(k)}}(z_k)  \right)
\end{split}
\ee
where $\sigma$ is a permutation of $N$ elements, and $(-1)^{| \sigma |}$ is the signature of the permutation. Thus we see that $\Phi_{Z}\Delta_{Za}$ ( $\Phi_{W}\Delta_{Wa}$ ) is equivalent to the sum of replacements of a single orbital $\psi_{n,m} \rightarrow \psi_{n+a,m-a}$. In the diagrammatic notation, this amounts to the sum of all moves taking a $Z$ (a $W$) $a$ slots ``up and to the left'' in the $\Lambda$-level diagram. If we take $\Phi^a_i$ to mean the Slater determinant $\Phi$ where the orbital $\psi_{n,m}$ in row $i$ of the determinant matrix has been replaced by $\psi_{n+a,m-a}$, then
\be{}
\left( \sum_{i=1}^N (\Phi_{Z})^a_i (\Phi_{W}) +
\sum_{j=1}^M (\Phi_{Z}) (\Phi_{W})^a_j \right) J =
\Phi_{Z} \Phi_{W} \left(\Delta_{Za} + \Delta_{Wa} \right) J
=0 \label{gen-trans-comp}
\ee
due to (\ref{jas-gen-trans}). This means that we can use generalised translation invariance to find dependence relations as we did for simple states. Notice that (\ref{gen-trans-comp}) is \emph{not} equivalent to having $\Delta_{Za} + \Delta_{Wa}$ to the left of the Slater determinants. For compact states we generally do not have
\be{}
\left(\Delta_{Za} + \Delta_{Wa} \right) \Phi_{Z} \Phi_{W} J = 0 \quad a>1.
\ee

We will now see that some care needs to be taken when applying this, since some resulting terms may not be compact. Consider the one-component CF candidate with Slater determinant
\be{}
\Phi_{W} = 
\begin{vmatrix}
 (w_1^0 \p^0_{w_1} ) \, (w_2^0 \p^0_{w_2} ) \\
(w_1^1 \p^0_{w_1} ) \, (w_2^1 \p^0_{w_2} ) \\
\end{vmatrix}
\
\ee
i.e. two CFs in the lowest $\Lambda$-level. The action of this determinant on $\Delta_W$ is
\be{}
\Phi_{W} \Delta_W = 
\begin{vmatrix}
 (w_1^0 \p^1_{w_1} ) \, (w_2^0 \p^1_{w_2} ) \\
(w_1^1 \p^0_{w_1} ) \, (w_2^1 \p^0_{w_2} ) \\
\end{vmatrix} +
\begin{vmatrix}
 (w_1^0 \p^0_{w_1} ) \, (w_2^0 \p^0_{w_2} ) \\
(w_1^1 \p^1_{w_1} ) \, (w_2^1 \p^1_{w_2} ) \\
\end{vmatrix}
\
\ee
and since the two resulting determinants are not compact, they are not even CF Slater determinants: the CF orbitals for a non-compact state have the general form (\ref{spwf2}), which these determinants clearly do not. However, this will not prevent us from applying translation invariance. We will simply include these non-CF contributions as unknown functions $\chi_i$ on the same footing as the compact states when writing down the linear relations, and attempt to eliminate them when performing the row reduction to independent wave functions. In the worst case scenario, all equations will still contain these non-CF states after row reduction, and we will have learned nothing about dependencies within the set of compact candidates. However, the results presented in \ref{sec:comb-res-gen-compact} indicate that, at least for the lowest sub-band of compact CF states, this never occurs, i.e. we are in fact able to find \emph{all} dependencies, and thus reduce the set of candidates to a proper basis, as was the case for simple states.

\subsection{Blocks}

From the form of Eq. (\ref{sps-compact-1}), it follows that the appearance of non-zero powers of the coordinates $z_i$, $w_j$ in the CF orbitals does not affect which terms in the Jastrow factor that survive the differentiation imposed by the Slater determinants. Therefore, the classification of blocks is also possible for compact states, and is simply achieved by counting derivatives as for simple states. For instance, the Slater determinant pair represented by the diagram in (\ref{diagram-example}) splits into two blocks: the first block $X_1$ contains $\Lambda$-levels 0 and 1, and the second $X_2$ contains $\Lambda$-levels 2 through 5. Notice that $X_2$ is in fact a simple block, and that we may apply reflection symmetry to $X_2$ to immediately acquire the relation
\be{}
\left(
\begin{array}{ccccccccc}
 & 5 & - & - & - & - & - & - & -\\
 & 4 & & W & - & - & - & - & -\\
 & 3 & & & Z & - & - & - & -\\
n & 2 &  & & & P & - & - & -\\
 & 1 & & &  & & - & - & -\\
 & 0 & &  &  &  & & W & W\\
 &  & -5 & -4 & -3 & -2 & -1 & 0 & 1 \\
 &  &  &  & m
 \end{array} \right) = (-1) \cdot \left(
\begin{array}{ccccccccc}
 & 5 & - & - & - & - & - & - & -\\
 & 4 & & Z & - & - & - & - & -\\
 & 3 & & & W & - & - & - & -\\
n & 2 &  & & & P & - & - & -\\
 & 1 & & &  & & - & - & -\\
 & 0 & &  &  &  & & W & W\\
 &  & -5 & -4 & -3 & -2 & -1 & 0 & 1 \\
 &  &  &  & m
 \end{array} \right)
\ee
In general, we may apply reflection symmetry to all simple blocks of a candidate, even though the full state is not simple. This is as far as we have been able to pursue the reflection concept for compact states.

Now we generalise permutations of blocks to compact states. It should be clear that any block spanning more than one $\Lambda$-level (more than a single $Z$ or $W$) necessarily ends on an empty level. However, a block may begin on a level containing multiple occupancies, and in particular, the lowest $\Lambda$-level can contain any number of particles. Permuting two blocks may then violate the second condition for compactness. If this is the case, then the permutation of this block with any other block will result in a non-CF wave function, with elements not corresponding to the correct expressions for the orbitals. However, we may use this fact to get rid of some of the variables $\chi_i$ that result from the translation invariance equations. To illustrate this, consider the action of the block
\be{}
\begin{array}{ccccccccc}
 & 4 & - & - & - & - & - & -\\
 & 3 & & Z & - & - & - & -\\
n & 2 & & & W & - & - & -\\
 & 1 & &  & & - & - & -\\
 & 0 &  &  &  & & P & W\\
 & & -4 & -3 & -2 & -1 & 0 & 1 \\
 &  &  &  & m
 \end{array}
\ee
on $\Delta_{W2}$. Among other terms, this will produce
\be{}
\begin{array}{ccccccccc}
 & 4 & - & - & - & - & - & -\\
 & 3 & & Z & - & - & - & -\\
n & 2 & & & W & W & - & -\\
 & 1 & &  & & - & - & -\\
 & 0 &  &  &  & & P & -\\
 & & -4 & -3 & -2 & -1 & 0 & 1 \\
 &  &  &  & m
 \end{array} \label{seems-not-compact}
\ee
but with the second W in $\Lambda$-level 2 given by $w_i \p_{w_i}^2$, which is not the correct expression for the $\psi_{2,-1}$ orbital in this non-compact case. However, we realize that permuting the two blocks of the compact candidate
\be{}
\begin{array}{ccccccccc}
 & 4 & - & - & - & - & - & -\\
 & 3 & & P & - & - & - & -\\
n & 2 & & & - & - & - & -\\
 & 1 & &  & & Z & - & -\\
 & 0 &  &  &  & & W & W\\
 & & -4 & -3 & -2 & -1 & 0 & 1 \\
 &  &  &  & m
 \end{array}
\ee
will give exactly (\ref{seems-not-compact}) apart from a minus sign given by Eq. (\ref{commutation-rule}). We conclude that some terms resulting from applying $\Delta_{Zn}$ or $\Delta_{Wn}$ may at first not seem compact, but turn out to be just that by carefully considering permutations of blocks. It is this fact that allows us to eliminate many unknown functions $\chi_i$ from the set of dependence relations, which in turn gives us more information about the dependencies between the compact candidates.

\subsection{Combined results}
\label{sec:comb-res-gen-compact}

The modifications described above are reflected in modifications to the algorithm described in section \ref{sec:comb-res-simple}. To obtain results for the lowest part of the yrast spectrum, we will
consider the CF candidates that, at a given $N,M,L$, minimize the CF cyclotron energy
\be{}
K = \sum_{i=1}^{N+M} n_i
\ee
where $n_i$ are the $\Lambda$-level quantum numbers of the occupied CF orbitals. We immediately notice that the application of Slater determinants on $\Delta_{Za} + \Delta_{Wa}$ raises $K$ and lowers $L$ by $a$ units, and that permutation of blocks and reflection of simple blocks leave $K$ invariant. Remember also that the action of a block with $N+M$ particles on $\Delta_{Za} + \Delta_{Wa}$ trivially vanishes if $a \geq N+M$, because too many derivatives will act on the Jastrow factor.

The algorithm can be summarized in the following steps:
\begin{enumerate}
\item Construct all compact CF candidates at $N,M,L$ with minimal CF cyclotron energy, $K_{min}$. This set of candidates is named $\left\{ \Psi_{CF}^{K_{min}} \right\}$.
\item Construct dependence relations by applying all combinations of permutation and reflection of blocks (reflection only for simple blocks). Use the dependence relations to reduce the set $\left\{ \Psi_{CF}^{K_{min}} \right\}$.
\item For all $\Delta L = 1,\ldots, N+M-1$, do:
\begin{enumerate}
\item Construct all compact CF candidates at $N,M,L+\Delta L$ with CF cyclotron energy $K = K_{min} - \Delta L$.
\item Construct additional dependence relations using the generalized translation invariance condition (\ref{gen-trans-comp}) with $a=\Delta L$, applied to all the blocks of the states in step 3(a). Simplify the new dependence relations using those found in step 2. Generally, some dependence relations will simplify to $0=0$, and some will contain unknown functions $\chi_i$.
\end{enumerate}
\item Using Gaussian elimination or similar methods, eliminate as many of the $\chi_i$ as possible.
\item Use the dependence relations not containing any $\chi_i$ to reduce $\left\{ \Psi_{CF}^{K_{min}} \right\}$ even further.
\end{enumerate}

\begin{figure}
\includegraphics[width=\columnwidth]{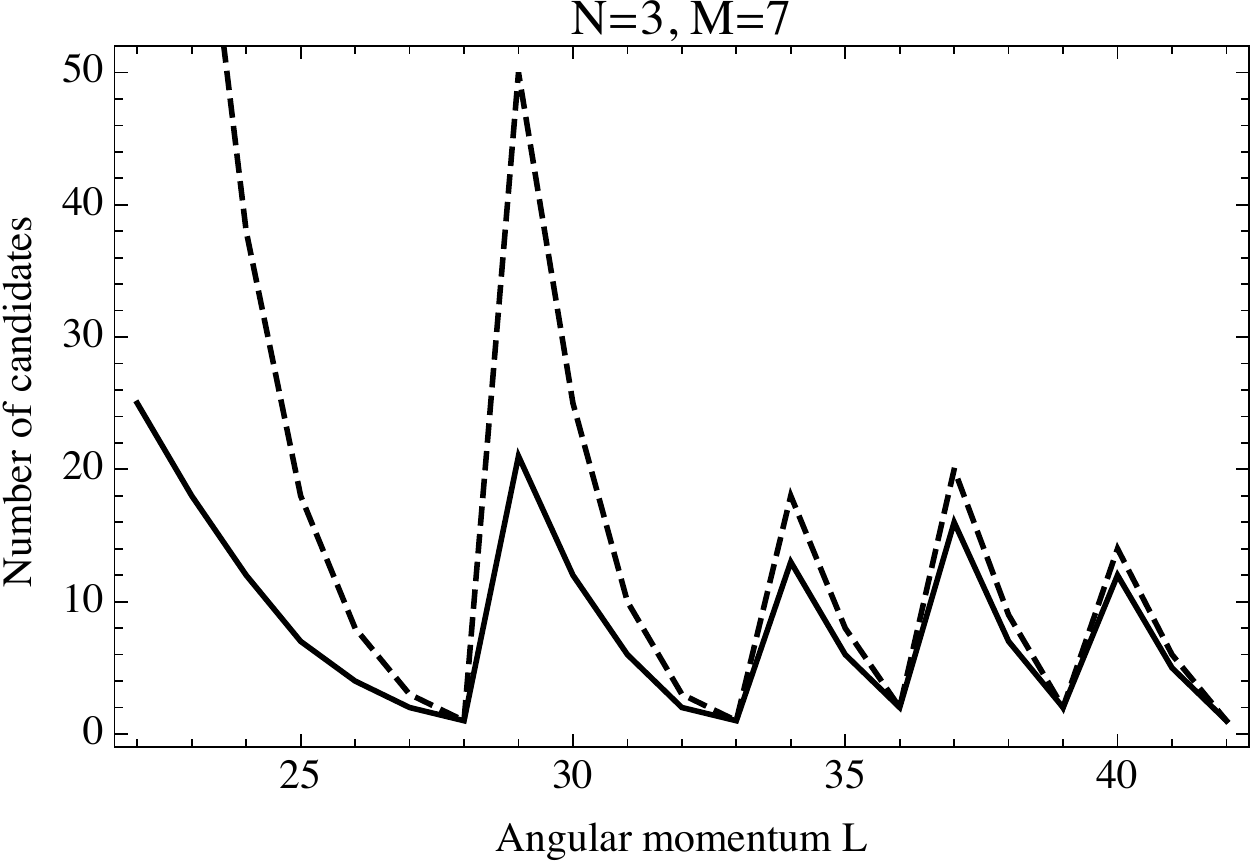}
\caption{The number of compact candidates before and after exploiting dependence relations, for $N=3$, $M=7$. The dashed line shows the naively independent states, while the solid line shows the remaining candidates after removing dependencies. For $42 \leq L \leq L_{221} = 69$ the two lines agree, i.e. the candidates are linearly independent to begin with.
\label{fig:compact}}
\end{figure}

Again we compare with a brute-force calculation of the number of linearly independent states. Surprisingly, we find that $\left\{ \Psi_{CF}^{K_{min}} \right\}$ is fully reduced in all cases, i.e. that the reduced $\left\{ \Psi_{CF}^{K_{min}} \right\}$ is a linearly independent set of wave functions, and thus constitutes a basis for the CF candidates in the lowest sub-band $K=K_{min}$. We have checked this for all 2072 combinations of $N,M,L$ where $N+M=2,\ldots,12$; $N=0,\ldots,\left \lfloor (N+M)/2 \right \rfloor$; $L=N \cdot M +1,\ldots, L_{221}$. An example of the numbers of candidates before and after exploiting dependence relations is seen in FIG. \ref{fig:compact}, for 3+7 particles. A clear pattern is visible, where the two lines agree in the cusps at certain values of $L=L_i$. The values of $L_i$ are characterized by having a unique state with minimal CF cyclotron energy $K_i$ such that, at $L=L_i+1$, there exist at least one state with the \emph{same} $K=K_i$. This happens when it is not possible to reduce $K$ by increasing $L$: at $L$ not equal to any of the $L_i$, $K$ strictly decreases with increasing $L$.

%
%

For higher sub-bands $K>K_{min}$, we find some, but not all, linear dependencies. Specifically, there exist dependencies between linear combinations of states in one band, and linear combinations of states in \emph{lower} bands. As our methods only produce dependence relations between states with equal $K$, we do not capture inter-band relations. We do, however, capture all intra-band dependencies. We expect that one would need to explicitly treat the non-compact contributions $\chi_i$ in order to fully understand inter-band dependencies. This is left for future work on the subject.


\section{Electronic CF states}
\label{sec:fermi}
We briefly comment on the application of this work's findings to CF states for the standard quantum Hall effect -- one-component electron systems in magnetic fields. We maintain the disk geometry and the projection method used so far, and comment on other geometries and projection methods later. The CF candidate wave functions in the disk are given by
\be{}
\Psi_{CF} = \mathcal{P}_{LLL} \left( \Phi J^{2p} \right)
\ee
with the usual interpretation that each electron captures an even number, $2p$, of vortices to form composite fermions. 
In contrast to the bosonic case, an even power of Jastrow factors, $J^2$ in the simplest case, does not only contain terms with one set of exponents like $J^1$ does. Therefore, there does not seem to be a straightforward way to identify blocks in a manner similar to what we have presented in this work. The largest allowed derivative in the Slater determinant is also higher now, so even if only the $m=-n$ CF orbitals are occupied, there will be more $H$'s than $P$'s in the string letter description, and we have not been able to give meaning to the reflection operation in this case.

For bosons, the result of acting on a (compact) CF candidate with the translation operator $\Delta_{Z}$ could be written as a sum of modified determinants acting on the unmodified Jastrow factor because $J$ itself is translation invariant. This is no less true for $J^{2p}$, and since the form of the Slater determinants is the same as before, translation invariance will reveal linear dependencies also for electronic CF wave functions, as long as they are compact. However, generalised translation invariance does not hold for $2p>1$:
\be{}
\Delta_{Za} J^{2p} \neq 0 \quad a>1,\ 2p > 1
\ee
To see how $\Delta_Z$ acts on a Slater determinant, we write
\be{}
\Delta_Z \mathcal{P}_{LLL} \left( \Phi J^{2p} \right)
= \mathcal{P}_{LLL} \left( \sum_{i=1}^N \zbar_i \Phi J^{2p} \right)
\ee
Before projection, $z$ and $\zbar$ commute, so we can evaluate the unprojected part of the right hand side of the equation. We need the following property of associated Laguerre polynomials:
\be{}
x L_n^m(x) = (n+m) L_n^{m-1} (x) - (n+1)L_{n+1}^{m-1} (x)
\ee
From this we easily find
\be{}
\begin{split}
\zbar \psi_{n,m} (z,\zbar) & = \zbar L_n^m(z \zbar) z^{m} \\
& = z^{m-1} z \zbar L_n^m (z \zbar) \\
& = (n+m) L_n^{m-1} (z \zbar) z^{m-1} - (n+1) L_{n+1}^{m-1}(z \zbar) z^{m-1} \\
& = (n+m) \psi_{n,m-1}(z,\zbar) - (n+1)\psi_{n+1,m-1}(z,\zbar)
\end{split} \label{trans-general}
\ee
where we have ignored normalization of $\psi_{n,m}$ and a factor $1/2$ from the argument of $L_n^{m}$. The interpretation is that $\Delta_Z \mathcal{P}_{LLL} \left( \Phi J^{2p} \right)$ is equal to a linear combination of CF candidates, where each term has one CF moved either to the left, or up and to the left in the $\Lambda$-level diagram. Now, if $\Phi$ is compact, then
\be{}
\Delta_Z \mathcal{P}_{LLL} \left( \Phi J^{2p} \right) = 0
\ee
We may use this to find linear dependencies between compact and non-compact states. For instance, the $\nu = 1/3$ state for $N=3$ has the CF-orbitals $(n,m)=(0,0),(0,1),(0,2)$ occupied, and applying $\Delta_Z$ to this state gives
\be{}
\mathcal{P}_{LLL} \left( \left(-\Phi_1 -\Phi_2 -\Phi_3 \right) J^2 \right) = 0
\ee
where $\Phi_1,\Phi_2,\Phi_3$ are the three possible composite fermion particle-hole excitations with $\Delta K = 1$, $ \Delta L = -1$.


\section{Discussion and outlook}
 \label{sec:concl}
The main result of this paper is a comprehensive algorithm that reduces the set of seemingly different, low-lying compact composite fermion candidates at given $N,M,L$ to a basis set of linearly independent states, \emph{prior} to LLL projection. This may amount to reducing the number of states by as much as orders of magnitude. The algorithm is based on exploiting translation invariance, and invariance of the final wave function under certain permutations of the Slater determinant occupation patterns, to identify the hidden linear dependencies. We presented the derivation in the context of so-called simple states for two-species bosons, and then outlined the pertinent modifications for general compact, one- and two-species bosons. In all cases, we found complete reduction to a basis when the lowest sub-band was considered.

At this time we have not yet been able to fully prove that generalised translation invariance (possibly helped by reflection and block permutation) is sufficient to produce a basis of simple candidates for general $N$, $M$, $L$. However we have found a recursive formula for the number of linearly independent states \cite{lia-meyer-future}, only involving $N,M,L$, and work is underway to complete the proof that generalised translation invariance explains all the dependencies between the simple CF candidates. We hope that this will lead to a revised algorithm that directly produces bases for the space spanned by the simple states.

We also commented on the case of fermions which is more complicated, since many of our results mathematically rely on the single-flux attachment of the boson case. A possible way around this might be to study fermionic CF-type wave functions of the form $\psi_F = J\cdot\psi_B$, where $J$ is a Jastrow factor. In other words, instead of the usual double flux attachment for fermions, leave one of the Jastrow factors ``outside'' the projection. Essentially this would amount to a somewhat different LLL projection method, and one would have to test numerically if it produces good wave functions. This is left for future study.
 
Although different choices of projection tend not to matter much when it comes to overlaps, energies etc. of CF trial states, they do produce somewhat different wave functions. There is thus reason to believe that using, e.g. Jain-Kamilla projection instead of ``brute force'' projection, the analytically exact linear dependencies of this paper might only be near-exact. However, it is possible that exact identities similar to those presented in this paper exist also for other projection methods, and that the ideas presented here may aid in identifying them. This, too, is left for future study. Similarly, it would be of interest to re-examine the issue in different geometries, in particular on the sphere.

Another obvious question is whether there is some more direct, qualitative or intuitive way of predicting the linear dependencies from the CF Slater determinants -- complementary to our rather mathematical algorithmic approach. We have not come up with a good answer so far. Indeed, some of the dependencies are so mathematically subtle that they may not have a simple, qualitative explanation within the CF phenomenology.


\section*{Acknowledgement}
\noindent
We would like to thank Jainendra Jain and Ajit Balram for inspiring discussions. This work was financially supported by the Research Council of Norway.
\




\appendix

\section{Proof of block permutation invariance}
\label{sec:app}

This appendix details the proof that permuting two blocks in a CF polynomial will leave the polynomial invariant up to a sign as stated earlier (\ref{commutation-rule}).
We consider a string with $K$ blocks.  The $j$'th block has $\nu_j$ $Z$'s and $\mu_j$ $W$'s.  The total number of $Z$($W$) is $N$($M$) as always.  Define the variables
\be{}
x_i=z_i\ \forall\ 0<i\leq N,\ x_i=w_{i-N}\ \forall\  N<i\leq N+M
\ee{}
and exponents
\be{}
\set{a_i\ |\ 0<i\leq N},\ \set{a_i\ |\ N<i\leq N+M}
\ee{}
such that the first(second) set contains the exponents of the $z$($w$) differentiation operators in increasing order.  We can now use the symmetric group of $n$ elements, $S_n$ to write the CF polynomial as
\be{}
\Psi=\sum_{\sigma\in S_N\oplus S_M}\sum_{\rho\in S_{N+M}}(-1)^{|\sigma|+|\rho|}\prod_{i=1}^{N+M}\frac{\partial^{a_i}}{\partial x_{\sigma_i}^{a_i}}\prod_{k=1}^{N+M}x^{k-1}_{\rho_k}.
\ee{}
The next step is to factorize each permutation into one permutation distributing the variables in the different blocks and one permutation permuting within the blocks.  We define
\be{}
S_\nu=\bigoplus_{j=1}^K S_{\nu_j},\ S_\mu=\bigoplus_{j=1}^K S_{\mu_j},\ S_{\nu+\mu}=\bigoplus_{j=1}^K S_{\nu_j+\mu_j}
\ee{}
and the quotient (not a group)
\be{}
S_/=S_N/S_\nu\oplus S_M/S_\mu=\set{\min(\set{\sigma\circ\rho|\sigma\in S_\nu\oplus S_\mu})|\rho\in S_N\oplus S_M},
\ee{}
where $\min()$ takes the lexicographically smallest element.  In words, for each distributition of $\set{1,...,N+M}$ into $2K$ parts of lengths $(\nu_1,..., \nu_K,\mu_1,...,\mu_K)$, $S_/$ contains one permutation with this distribution and the permutation is sorted in ascending order within each part.  We can now factorize the determinant permutations as
\be{}
\Psi=\sum_{\sigma\in S_/}\sum_{\rho\in S_\nu\oplus S_\mu}\sum_{\tau\in S_{N+M}}(-1)^{|\sigma|+|\rho|+|\tau|}\prod_{i=1}^{N+M}\frac{\partial^{a_i}}{\partial x_{(\rho\circ\sigma)_i}^{a_i}}\prod_{k=1}^{N+M}x^{k-1}_{\tau_k}.
\ee{}
We use that $\tau\in S_{N+M}$ must distribute $\nu_j$ $z$'s and $\mu_j$ $w$'s in the $j$'th block for the contribution to be non-zero.  Let us define the permutation 
\be{}
\sigma'=(1,...,\nu_1, N+1,..., N+\mu_1, \nu_1+1,.\ .\ .\ , M)
\ee{}
which is the lexicographically smallest such permutation.  One then arrives at the factorization
\be{}
\tau=\rho\circ\sigma'\circ\sigma,\ \rho\in S_{\nu+\mu},\ \sigma\in S_/.
\ee{}
Also, since we require that the indices distributed in the differentiation blocks are the same as the indices distributed in the variables, the $\sigma$ permutations must be the same.  This allows us to write
\be{}
\Psi=\sum_{\sigma\in S_/}\sum_{\rho\in S_\nu\oplus S_\mu}\sum_{\tau\in S_{\nu+\mu}}(-1)^{2|\sigma|+|\rho|+|\tau|+|\sigma|'}\prod_{i=1}^{N+M}\frac{\partial^{a_i}}{\partial x_{(\rho\circ\sigma)_i}^{a_i}}\prod_{k=1}^{N+M}x^{k-1}_{(\tau\circ\sigma'\circ\sigma)_k}.
\ee{}
Some further notation is needed to reduce this to an explicitly block order independent expression.  We introduce the subgroups
\be{}
(S_\alpha)_j=\bigoplus_{i=1}^{j-1}\id_{\alpha_i}\oplus S_{\alpha_j}\oplus\bigoplus_{i=j+1}^{K}\id_{\alpha_i}\subset S_\alpha,\ \alpha\in\set{\nu, \mu, \nu+\mu},\ j\in\set{1,...,K}.
\ee{}
Also, for $j\in\set{1,...,K}$ we define
\be{}
n_{ji}=\left\{ \begin{array}{ll}
i+\sum_{h=1}^{j-1}\nu_h & \forall\ 0<i\leq \nu_j\\
N+i-\nu_j+\sum_{h=1}^{j-1}\mu_h & \forall\ \nu_j<i\leq\nu_j+\mu_j \end{array}\right.
\ee{}
such that $n_{ji}$ is the $i$'th index of the $j$'th block.  We can now write
\be{}
\Psi=(-1)^{|\sigma'|}\sum_{\sigma\in S_/}\prod_{j=1}^K\sum_{\rho\in (S_\nu)_j\oplus (S_\mu)_j}\sum_{\tau\in (S_{\nu+\mu})_j}(-1)^{|\rho|+|\tau|}\prod_{i=1}^{\nu_j+\mu_j}\frac{\partial^{a_{n_{ji}}}}{\partial x_{(\rho\circ\sigma)_{n_{ji}}}^{a_{n_{ji}}}}\prod_{k=1}^{\nu_j+\mu_j}x^{n_{jk}-1}_{(\tau\circ\sigma'\circ\sigma)_{n_{jk}}}.
\ee{}
All the differentiation operators in the $j$'th block are of order at least $n_{j1}$.  We can carry out this differentiation in each block to obtain
\be{}
\Psi=(-1)^{|\sigma'|}\sum_{\sigma\in S_/}\prod_{j=1}^K\sum_{\rho,\tau}(-1)^{|\rho|+|\tau|}\prod_{i=1}^{\nu_j+\mu_j}\frac{\partial^{a_{n_{ji}}-n_{j1}}}{\partial x_{(\rho\circ\sigma)_{n_{ji}}}^{a_{n_{ji}}-n_{j1}}}\prod_{k=1}^{\nu_j+\mu_j}x^{k-1}_{(\tau\circ\sigma'\circ\sigma)_{n_{jk}}}\frac{(n_{jk}-1)!}{(k-1)!}.
\ee{}
The product over $j$ can be carried out for the numerical factor and we get
\be{}
\Psi=(-1)^{|\sigma'|}C\sum_{\sigma\in S_/}\prod_{j=1}^K\sum_{\rho\in (S_\nu)_j\oplus (S_\mu)_j}\sum_{\tau\in (S_{\nu+\mu})_j}(-1)^{|\rho|+|\tau|}\prod_{i=1}^{\nu_j+\mu_j}\frac{\partial^{a_{n_{ji}}-n_{j1}}}{\partial x_{(\rho\circ\sigma)_{n_{ji}}}^{a_{n_{ji}}-n_{j1}}}\prod_{k=1}^{\nu_j+\mu_j}x^{k-1}_{(\tau\circ\sigma'\circ\sigma)_{n_{jk}}},
\ee{}
where
\be{}
C=\frac{\prod_{i=0}^{N+M-1}i!}{\prod_{j=1}^K\prod_{i=0}^{\nu_j+\mu_j-1} i!}.
\ee{}
This factor is independent of the order of blocks, and the same is true for the 
reduced differentiation exponents, $a_{n_{ji}}-n_{j1}$.  The only block order dependent part of the polynomial is the sign $(-1)^{|\sigma'|}$.  Finally, let us deduce the sign rule.

 It is sufficient to consider a wave function with only two blocks $X_1 \circ X_2$. The two different orders of blocks will have:
\be{}\label{prim1}
\sigma_1' : z_1...z_N w_1...w_M \rightarrow z_1...z_{\nu_1} w_1...w_{\mu_1} z_{\nu_1+1}...z_N w_{\mu_1+1}...w_M
\ee{}
and
\be{}\label{prim2}
\sigma_2' : z_1...z_N w_1...w_M \rightarrow z_1...z_{\nu_2}w_1...w_{\mu_2}z_{\nu_2+1}...z_N w_{\mu_2+1}...w_M
\ee{}
respectively. The signatures are
\be{}
(-1)^{|\sigma_1'|} = (-1)^{\nu_2 \mu_1},\quad (-1)^{|\sigma_2'|} = (-1)^{\nu_1 \mu_2}
\ee
The relative signature of the two $\sigma'$ is the product of the signatures:
\be{}
(-1)^{|\sigma_1'|+|\sigma_2'|}=(-1)^{\nu_2\mu_1+\nu_1\mu_2}.
\ee{}
which was stated in Eq. (\ref{commutation-rule}).

\end{document}